\documentclass[aps,pra,twocolumn,reprint,superscriptaddress]{revtex4-2}

\usepackage{graphicx}
\usepackage[english]{babel}
\usepackage{amsmath, amssymb, amsfonts}
\usepackage{bm}
\usepackage{color}
\usepackage[normalem]{ulem}

\begin{document}
\title{Tunable Feshbach resonances in collisions of ultracold molecules in $^2\Sigma$ states \\
with alkali-metal atoms}

\author{Robert C. Bird}
\affiliation{Joint Quantum Centre (JQC) Durham-Newcastle, Department of Chemistry, Durham University, South Road, Durham, DH1 3LE, United Kingdom.}
\author{Michael R. Tarbutt}
\affiliation{Centre for Cold Matter, Blackett Laboratory, Imperial College London, Prince Consort Road, London SW7 2AZ UK}
\author{Jeremy M. Hutson}
\email{j.m.hutson@durham.ac.uk}
\affiliation{Joint Quantum Centre (JQC) Durham-Newcastle, Department of Chemistry, Durham University, South Road, Durham, DH1 3LE, United Kingdom.}

\date{\today}

\begin{abstract}
	We consider the magnetically tunable Feshbach resonances that may exist in ultracold mixtures of molecules in $^2\Sigma$ states and alkali-metal atoms. We focus on Rb+CaF as a prototype system. There are likely to be Feshbach resonances analogous to those between pairs of alkali-metal atoms. We investigate the patterns of near-threshold states and the resonances that they cause, using coupled-channel calculations of the bound states and low-energy scattering on model interaction potentials. We explore the dependence of the properties on as-yet-unknown potential parameters. There is a high probability that resonances will exist at magnetic fields below 1000~G, and that these will be broad enough to control collisions and form triatomic molecules by magnetoassociation. We consider the effect of CaF rotation and anisotropy of the interaction potential, and conclude that they may produce additional resonances but should not affect the existence of rotation-free resonances.
\end{abstract}

\maketitle

\section{Introduction}

Ultracold molecules have many applications that are now emerging, ranging from quantum simulation \cite{Gorshkov:2011, Blackmore:2019}, quantum computing \cite{DeMille:2002, Holland:2023, Bao:2023}, the study of novel quantum phases \cite{Yan:2013, Li:2023}, and tests of fundamental physics \cite{Fitch:2021, Anderegg:2023, Leung:2023}. Key to most of these applications are polar molecules, which can have long-range anisotropic interactions resulting from their permanent dipoles. Many such molecules have been produced at microkelvin temperatures by association of pairs of alkali-metal atoms, followed by laser transfer to the vibrational ground state \cite{Ni:KRb:2008, Takekoshi:RbCs:2014, Molony:RbCs:2014, Park:NaK:2015, Guo:NaRb:2016, Rvachov:2017, Seesselberg:2018,
Yang:K_NaK:2019, Voges:NaK:2020, Cairncross:2021}. Another class of molecules, exemplified by CaF and SrF, have been cooled directly by magneto-optical trapping followed by sub-Doppler laser cooling \cite{Truppe:MOT:2017, McCarron:2018, Anderegg:2018, Cheuk:2018, Caldwell:2019, Ding:2020}.

Elastic and inelastic collisions are at the heart of ultracold physics. For ultracold atoms, it is often possible to control ultracold collisions by adjusting an applied magnetic field close to a zero-energy Feshbach resonance \cite{Chin:RMP:2010}. Such a resonance occurs whenever a molecular bound state can be tuned across a scattering threshold as a function of applied field. The s-wave scattering length then passes through a pole as a function of field, allowing the effective interaction strength to be tuned to any desired value. This control has been applied in many areas of ultracold physics, including condensate collapse \cite{Roberts:collapse:2001}, soliton creation \cite{Strecker:soliton:2003}, Efimov physics \cite{Naidon:2017} and investigations of the BCS-BEC crossover in degenerate Fermi gases \cite{Bloch:2012}. Feshbach resonances are also used for magnetoassociation, in which pairs of ultracold atoms are converted to weakly bound diatomic molecules by sweeping a magnetic field across the resonance \cite{Hutson:IRPC:2006, Kohler:RMP:2006}.

Much new physics will become accessible when atom-molecule collisions can be controlled with tunable Feshbach resonances. Control of the s-wave scattering length may allow sympathetic cooling of molecules to quantum degeneracy, and the formation of atom-molecule mixtures with novel properties. It may also be possible to form polyatomic molecules by magnetoassociation. Feshbach resonances have now been observed in collisions between ultracold $^{40}$K atoms and $^{23}$Na$^{40}$K molecules in singlet states \cite{Yang:K_NaK:2019, Wang:K_NaK:2021, Su:elastic:2022, Yang:rf-association:2022, Yang:magnetoassociation:2022} and between $^{23}$Na atoms and $^6$Li$^{23}$Na molecules in triplet states \cite{Son:Na_NaLi:2022}. These systems have also been investigated theoretically \cite{Wang:K_NaK:2021, Frye:triatomic-complexes:2021, Son:Na_NaLi:2022, Frye:long-range:2023}. Resonances have not yet been observed in collisions of laser-cooled molecules such as CaF and SrF, with $^2\Sigma$ ground states, but we have recently succeeded in making ultracold mixtures of CaF molecules and Rb atoms, and studied their inelastic collisions in both magnetic traps \cite{Jurgilas:magnetic:2021} and magneto-optical traps \cite{Jurgilas:MOT:2021}. Several laser-coolable molecules have been cooled to 5 $\mu$K \cite{Cheuk:2018, Caldwell:2019} and confined in optical traps \cite{Anderegg:2018} and optical tweezers \cite{Anderegg:2019}, opening the way to experiments in controlled magnetic fields.

The purpose of the present paper is to investigate the resonances that are expected in collisions between molecules in $^2\Sigma$ states and alkali-metal atoms. These systems have strong similarities to pairs of alkali-metal atoms, particularly for the long-range states that are most likely to cause magnetically tunable Feshbach resonances. We show that there is a high probability that tunable Feshbach resonances will exist at magnetic fields below 1000~G, and that they will be broad enough to control collisions and form triatomic molecules by magnetoassociation. There are additional complications and additional resonances that arise from the rotational structure of the molecule and the anisotropy of the interaction potential, but we find that these are unlikely to affect the general features of the scattering. We focus on $^{87}$Rb+$^{40}$Ca$^{19}$F as a prototype system, but many of the features are transferable to other molecules such as SrF and other alkali-metal atoms. In the following we mostly omit isotopic masses and write Rb+CaF for $^{87}$Rb+$^{40}$Ca$^{19}$F.

The structure of the paper is as follows. Section \ref{sec:theory} describes the underlying theory, including monomer Hamiltonians, interaction potentials, and computational methods.  Section \ref{sec:crossings0} describes the near-threshold levels that can exist for Rb+CaF and the Feshbach resonances they can cause, using a simple model that omits rotational degrees of freedom. Section \ref{sec:rot} considers the effects of CaF rotation and potential anisotropy. Section \ref{sec:chaos} considers the possible effects of quantum chaos at short range. Section \ref{sec:conc} presents conclusions and offers perspectives for future work to take advantage of the resonances.

\section{Theory}
\label{sec:theory}

\subsection{Monomer Hamiltonians and levels}

\begin{figure*}
\begin{center}
\includegraphics[width=2.0\columnwidth]{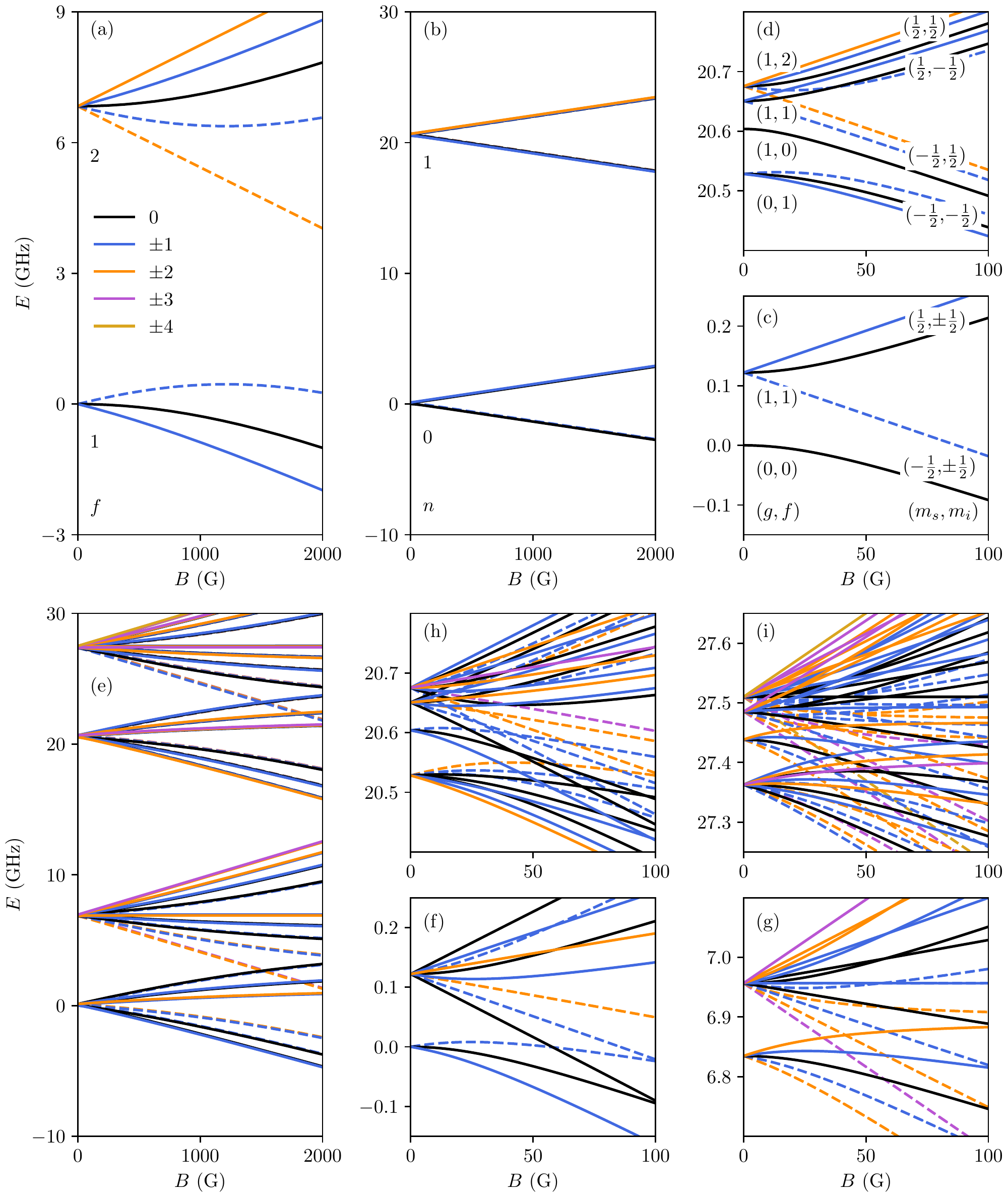}
\caption{Energies as a function of magnetic field for (a) $^{87}$Rb atom in ground $^2$S state; (b) Lowest two rotational levels of CaF, with expanded views of $n=0$ and 1 in (c) and (d), respectively; (e) Scattering thresholds of $^{87}$Rb+CaF, with expanded views of $(f_\textrm{Rb},n)=(1,0)$, (2,0), (1,1) and (2,1) in (f) and (g), (h) and (i), respectively. All level energies are shown relative to the ground state at zero field and are color-coded as shown in the legend according to $m_{f,\textrm{Rb}}$, $m_{f,\textrm{CaF}}$ or $M_F=m_{f,\textrm{Rb}}+m_{f,\textrm{CaF}}$, as appropriate; negative values are indicated by dashed lines.
\label{fig:levels}
}
\end{center}
\end{figure*}

The Hamiltonian of an alkali-metal atom A in its ground $^2$S state is
\begin{equation}
\hat{h}_\textrm{A} = \zeta_\textrm{A} \hat{\boldsymbol{i}}_\textrm{A}\cdot\hat{\boldsymbol{s}}_\textrm{A}
+ \left( g_{s,\textrm{A}} \hat{s}_{\textrm{A},z} + g_i\hat{i}_{\textrm{A},z} \right) \mu_\textrm{B} B,
\label{eq:ham-atom}
\end{equation}
where $\hat{\boldsymbol{s}}_\textrm{A}$ and $\hat{\boldsymbol{i}}_\textrm{A}$ are vector operators for the electron and nuclear spin, $\hat{s}_{\textrm{A},z}$ and $\hat{i}_{\textrm{A},z}$ are their components along the $z$ axis defined by the magnetic field $B$, $\zeta_\textrm{A}$ is the hyperfine coupling constant, and $g_{s,\textrm{A}}$ and $g_{i,\textrm{A}}$ are the g-factors for the electron and nuclear spins \footnote{In writing basis sets for pairs of atoms, it is necessary to distinguish between quantum numbers for the individual atoms and those for the pair. We adopt the widely used convention of using lower-case letters for the individual atoms and upper-case letters for the pair.}. The nuclear spins vary from 1 for $^6$Li to 9/2 for $^{40}$K, and the hyperfine splittings $A_\textrm{hfs} = \zeta_\textrm{A}(i_\textrm{A}+\frac{1}{2})/h$ vary from 228~MHz for $^6$Li to 9.19~GHz for $^{133}$Cs. We focus here on $^{87}$Rb, $i=3/2$ and $A_\textrm{hfs}\approx6.83$~GHz. The resulting levels are well known, but are shown in Fig.\ \ref{fig:levels}(a) for convenience. At zero field the levels are labeled by total angular momentum $f_\textrm{Rb}=1$ and 2. When a field is applied, each level splits into $2f_\textrm{Rb}+1$ sublevels, color-coded according to the projection $m_{f,\textrm{Rb}}$. At sufficiently high field, pairs of levels with $f_\textrm{Rb}=1$ and 2 but the same value of $m_{f,\textrm{Rb}}$ mix sufficiently that the levels are better described by $m_{s,\textrm{Rb}}$ and $m_{i,\textrm{Rb}}$ than by $f_\textrm{Rb}$. For $^{87}$Rb this transition is still incomplete at 2000~G, but it occurs at much lower fields for alkali-metal atoms with small hyperfine splittings, such as Li and Na.

The CaF or SrF molecule may be treated at different levels of complexity. The stable isotopes $^{40}$Ca, $^{88}$Sr, $^{86}$Sr and $^{84}$Sr all have zero nuclear spin, while $^{87}$Sr has $i=9/2$; only the spin-zero isotopes will be considered here. The simplest useful approximation is to neglect the molecular rotation, and in this case the molecular Hamiltonian $\hat{h}_\textrm{CaF}^{n=0}$ is the same as Eq.\ \ref{eq:ham-atom}, with $i_\textrm{F}=1/2$ for $^{19}$F in CaF. However, when rotation is included, several extra terms are needed. The ones important here are
\begin{equation}
\hat{h}_\mathrm{CaF}^{\mathrm{rfhf}} = b_0 \hat{\boldsymbol{n}}^2
+ \gamma\hat{\boldsymbol{s}}_\textrm{CaF} \cdot \hat{\boldsymbol{n}}
+ t\sqrt{6}T^2(C) \cdot T^2(\hat{\boldsymbol{i}}_\textrm{F},\hat{\boldsymbol{s}}_\textrm{CaF}),
\label{eq:hfhf-simp}
\end{equation}
where $\hat{\boldsymbol{n}}$ is the vector operator for the molecular rotation. The first term represents the rotational energy of a molecule in its vibrational ground state, treated as a rigid rotor. The second term represents the electron spin-rotation interaction, and the third accounts for the anisotropic interaction between electron and nuclear spins: $T^2(\hat{\boldsymbol{i}},\hat{\boldsymbol{s}})$ is the rank-2 spherical tensor formed from $\hat{\boldsymbol{i}}$ and $\hat{\boldsymbol{s}}$, and $T^2(C)$ is a spherical tensor whose components are the Racah-normalized spherical harmonics $C^2_q(\theta,\phi)$ involving the orientation of the molecular axis. Values of $b_0/h\approx10.3$~GHz, $\gamma/h\approx40$~MHz, $\zeta_\textrm{F}/h\approx120$~MHz and $t/h\approx14$~MHz are taken from ref.\ \cite{Childs:CaF:1981} \footnote{Ref.\ \cite{Childs:CaF:1981} uses the notation of Frosch and Foley \cite{Frosch:1952}, where our $b_0$, $\gamma$, $\zeta_\textrm{F}$ and $t$ are $B$, $\gamma$, $b+c/3$ and  $c/3$, respectively.}. A more complete version of Eq.\ \ref{eq:hfhf-simp}, including additional contributions of the order of kHz that are unimportant here, has been given in ref.\ \cite{Caldwell:long-coh:2020}.

The full Hamiltonian for CaF is $\hat{h}_\mathrm{CaF} = \hat{h}_\mathrm{CaF}^{n=0} + \hat{h}_\mathrm{CaF}^{\mathrm{rfhf}}$. The resulting level diagram is shown as a function of magnetic field in Fig.\ \ref{fig:levels}(b), with expanded views for $n=0$ and 1 in Figs.\ \ref{fig:levels}(c) and (d). There are only very small matrix elements that are off-diagonal in $n$, so the levels for $n=0$ are very similar to those of an alkali-metal atom with $i_\textrm{F}=1/2$. The hyperfine splitting is small, so $i_\textrm{F}$ and $s_\textrm{CaF}$ are mostly decoupled by 50~G. At higher field, the states are well described by $m_{s,\textrm{CaF}}$ and $m_{i,\textrm{F}}$.

In a rotating molecule at low field, $i_\textrm{F}$ and $s_\textrm{CaF}=1/2$ couple to give a resultant $g=0$ or 1, and $g$ couples to the rotational angular momentum $n$ to produce the total molecular angular momentum $f_\textrm{CaF}$. For $n=1$, there are zero-field levels with $f_\textrm{CaF}=0$, 1, 1, 2, as labeled on Fig.\ \ref{fig:levels}(d). The lower level with $f=1$ is predominantly $g=0$ and the remaining three are predominantly $g=1$. In a magnetic field, however, $i_\textrm{F}$, $s_\textrm{CaF}$ and $n$ are again mostly decoupled by 50~G; at higher fields, the states are better described by $m_{s,\textrm{CaF}}$, $m_{i,\textrm{F}}$ and $m_n$ than by $g$ and $f_\textrm{CaF}$. States of different $m_{s,\textrm{CaF}}$ are well separated; within the group for a particular value of $m_{s,\textrm{CaF}}$, there are 2 subgroups with $m_{i,\textrm{F}}=\pm\frac{1}{2}$, with splitting about $\zeta/2=60$~MHz, and each subgroup is further divided into states with different $m_n$, with adjacent states separated by about $\gamma/2=20$~MHz. The projection quantum numbers are not fully conserved, but these qualitative arguments help to understand the general patterns at high field.

\subsection{Calculations of bound states and scattering}
\label{sec:bound:scat}

The Hamiltonian for an alkali-metal atom interacting with a CaF molecule is
\begin{equation}
\hat{H} = \frac{\hbar^2}{2\mu}\left( -R^{-1} \frac{d^2}{dR^2} R + \frac{\hat{\boldsymbol{L}}^2}{R^2} \right)
+ \hat{h}_\textrm{A} + \hat{h}_\textrm{CaF}  + \hat{V}_\textrm{int},
\label{eq:ham-pair}
\end{equation}
where $R$ is the intermolecular distance, $\mu$ is the reduced mass, $\hat{\boldsymbol{L}}^2$ is the operator for relative rotation of the pair and $\hat{V}_\textrm{int}$ is the interaction operator described below.
We carry out calculations of both bound states and scattering using coupled-channel methods \cite{Arthurs:1960, Stoof:1988, Chin:RMP:2010}.
The total wavefunction is expanded as
\begin{equation} \Psi(R,\xi)
=R^{-1}\sum_j\Phi_j(\xi) \psi_{j}(R).
\label{eq:expand}
\end{equation}
Here $\{\Phi_j(\xi)\}$ is a set of basis functions that span all coordinates except $R$, including the relative rotation; these coordinates are collectively designated $\xi$. In the coupled-channel calculations describe in Sec.\ \ref{sec:crossings0}, $\xi$ includes only electron and nuclear spins. However, in more complete treatments, it may also include basis functions for overall rotation of the collision complex and rotation and vibration of CaF.

Substituting the expansion (\ref{eq:expand}) into the total Schr\"odinger equation produces a set of coupled differential equations that are solved by propagation with respect to the internuclear distance $R$. The coupled equations are identical for bound states and scattering, but the boundary conditions are different.

Scattering calculations are performed with the \textsc{molscat} package \cite{molscat:2019, mbf-github:2020}. Such calculations produce the scattering matrix $\boldsymbol{S}$, for a single value of the collision energy and magnetic field each time. The complex s-wave scattering length $a(k_0)$ is obtained from the diagonal element of $\boldsymbol{S}$ in the incoming channel, $S_{00}$,
\begin{equation}
a(k_0) = \frac{1}{ik_0} \left(\frac{1-S_{00}(k_0)}{1+S_{00}(k_0)}\right),
\end{equation}
where $k_0$ is the incoming wavenumber, related to the collision energy $E_\textrm{coll}$ by $E_\textrm{coll}=\hbar^2k_0^2/(2\mu)$. The scattering length $a(k_0)$ becomes constant at sufficiently low $E_\textrm{coll}$, with limiting value $a$. In the present work, s-wave scattering lengths are calculated at $E_\textrm{coll}/k_\textrm{B} = 10$~nK, which is low enough to neglect the dependence on $k_0$.

A zero-energy Feshbach resonance occurs where a bound state of the atom-molecule pair (triatomic molecule) crosses a scattering threshold as a function of applied field. At the lowest threshold, or in the absence of inelastic processes, the scattering length is real. Near a resonance, $a(B)$ passes through a pole, and is approximately
\begin{equation}
a(B) = a_\textrm{bg} \left( 1 - \frac{\Delta}{B-B_\textrm{res}}\right),
\label{eq:res}
\end{equation}
where $B_\textrm{res}$ is the position of the resonance, $\Delta$ is its width, and $a_\textrm{bg}$ is a slowly varying background scattering length. In the presence of inelastic processes, $a(B)$ is complex and the pole is replaced by an oscillation \cite{Hutson:res:2007}. \textsc{molscat} can converge on Feshbach resonances automatically and characterize them to obtain $B_\textrm{res}$, $\Delta$ and $a_\textrm{bg}$ (and the additional parameters needed in the presence of inelasticity) as described in ref.\ \citenum{Frye:resonance:2017}.

Coupled-channel bound-state calculations are performed using the packages \textsc{bound} and \textsc{field} \cite{bound+field:2019, mbf-github:2022}, which converge upon bound-state energies at fixed field, or bound-state fields at fixed energy, respectively. The methods used are described in ref.\ \citenum{Hutson:CPC:1994}.

In the present work, the coupled equations for both scattering and bound-state calculations are solved using the fixed-step symplectic log-derivative propagator of Manolopoulos and Gray \cite{MG:symplectic:1995} from $R_\textrm{min}=3\ a_0$ to $R_\textrm{mid}=15\ a_0$, with an interval size of $0.001\ a_0$, and the variable-step Airy propagator of Alexander and Manolopoulos \cite{Alexander:1987} between $R_\textrm{mid}$ and $R_\textrm{max}$, where $R_\textrm{max}=300\ a_0$ for \textsc{bound} and \textsc{field} and $3,000\ a_0$ for \textsc{molscat}.

\subsection{The interaction operator}

Rb($^2$S) and CaF($^2\Sigma$) interact to give two electronic surfaces of $^1$A$^\prime$ and $^3$A$^\prime$ symmetry. These are to some extent analogous to the singlet and triplet curves of alkali-metal dimers: the singlet surface is expected to be deep, and the triplet surface much shallower. The surfaces have not been characterized in any detail, either experimentally or theoretically, but both of them are expected to be strongly anisotropic at short range. We designate them $V^S(R,\theta)$, with $S=0$ for the singlet and $S=1$ for the triplet. Here $\theta$ is the angle between the CaF bond and the intermolecular axis in Jacobi coordinates. The interaction operator is
\begin{equation}
\hat{V}_\textrm{int} = V^0(R,\theta) \hat{\cal P}^0 + V^1(R,\theta) \hat{\cal P}^1 + \hat{V}^\textrm{d},
\end{equation}
where $\hat{\cal P}^0$ and $\hat{\cal P}^1$ are projection operators onto the singlet and triplet spin spaces, respectively, and $\hat{V}^\textrm{d}$ is a small electron spin-spin term described below.

The Feshbach resonances of interest here depend mostly on the properties of near-threshold states. These are bound by amounts comparable to the hyperfine and Zeeman splittings of Rb and CaF and (to a lesser extent) the low-lying rotational states of CaF. The most important states are those with binding energies less than about 30 GHz below their respective thresholds; this is considerably less than 0.1\% of the expected singlet well depth. The binding energies of these states are dependent mostly on long-range dispersion and induction forces, which are the same for the singlet and triplet surfaces. The leading term is of the form
\begin{equation}
V^S(R,\theta) = \left[ -C_6^{(0)} - C_6^{(2)} P_2(\cos\theta)\right] R^{-6},
\end{equation}
with $C_6^{(0)} \approx 3084\ E_\textrm{h}a_0^6$ \cite{Lim:sympa:2015}. For $C_6^{(2)}$ there is substantial cancelation between the dispersion and induction contributions; we estimate $C_6^{(2)} \approx 100(20)\ E_\textrm{h}a_0^6$. For Rb+CaF, the outer turning point at a binding energy of 30~GHz is near 30 $a_0$.

Potential terms that are the same for the singlet and triplet surfaces cannot cause couplings between orthogonal spin states. They are therefore unlikely to cause magnetically tunable Feshbach resonances. The most important interactions that mix different spin states are spin-exchange interactions, due to the difference between the singlet and triplet surfaces. Julienne \emph{et al.}\ \cite{Julienne:1997} have shown that, for a pair of atoms, spin-exchange interactions can cause nonadiabatic transitions between coupled channels at distances $R_\textrm{X}$ where the interaction approximately matches the asymptotic energy difference between the channels concerned. For $^{87}$Rb this occurs around $22\ a_0$ \cite{Julienne:1997}. The strength of the interaction is modulated by overall phases due to the short-range parts of the potentials for the channels concerned, and (if the long-range potentials are identical from $R_\textrm{X}$ to $\infty$) is smallest when the two channels have the same scattering length.

There is also a spin-spin term $\hat{V}^\textrm{d}$ in the interaction operator that results from magnetic dipole-dipole interactions between the electron spins on Rb and CaF, supplemented at short range by second-order spin-orbit terms that have the same overall dependence on spin coordinates. This term is important for heavy alkali-metal atoms such as Cs \cite{Berninger:Cs2:2013}, and may cause additional weak resonances in Rb+CaF as discussed below, but its effect is not considered in detail in the present work.

\subsection{Thresholds}

Figure \ref{fig:levels}(e) shows the scattering thresholds for $^{87}$Rb+CaF, which are simply sums of energies of Rb and CaF. Figures (f) to (i) show expanded views of each group. The thresholds are color-coded according to $M_F=m_{f,\textrm{Rb}}+m_{f,\textrm{CaF}}$, because this quantity is conserved in collisions if anisotropic terms in $V_\textrm{int}$ are neglected.

The importance of the thresholds lies in the fact that near-threshold levels lie approximately parallel to them, within well-defined energy intervals known as bins. This concept will be used extensively in discussing the patterns of near-threshold levels and the resulting resonances in the following sections.

\subsection{Near-threshold levels}

Each scattering threshold $j$ supports a series of levels of the collision complex that have binding energies $E^\textrm{b}_{j\eta}(B)$ below the threshold concerned. Here $\eta$ is a vibrational quantum number, defined so that the least-bound rotationless state below each threshold is labelled $\eta=-1$ and successively deeper levels are labeled $-2$, $-3$, etc. To a first approximation, the near-threshold levels retain the character of the threshold that supports them. Because of this, each level lies approximately parallel to the threshold that supports it and may be described in a single-channel approximation. There are nevertheless interactions between levels supported by different channels $j$, which cause $B$-dependent shifts and avoided crossings between levels. These interactions, and the strengths of the resulting avoided crossings, generally become larger as $|\eta|$ increases; these will be discussed below.

For a single-channel system with an asymptotic potential $-C_6R^{-6}$, the least-bound s-wave state (with $L=0$ and $\eta=-1$) lies within $\sim 36\bar{E}$ of threshold, where $\bar{E}=\hbar^2/(2\mu \bar{a}^2)$ and $\bar{a}$ is the mean scattering length of Gribakin and Flambaum \cite{Gribakin:1993}, $\bar{a}=(2\mu C_6/\hbar^2)^{1/4}\times 0.4779888\dots$. We refer to this energy interval as the top bin. The position of the bound state within this bin depends on the background scattering length $a_\textrm{bg}$ for the channel concerned, neglecting resonances (which themselves arise from couplings between channels). Each subsequent level $(\eta=-2$, $-3$, etc.) lies within its own bin, with successive bins becoming wider and bin boundaries at energies roughly proportional to $(|\eta|+\frac{1}{8}^3)$ \cite{LeRoy:1970, Gao:2000}. For Rb+CaF, $\bar{a}=67.3\ a_0$, $\bar{E}/h=11.4$~MHz, and the first 5 bin boundaries are at about 410, 2900, 9100, 21000 and 40000 MHz. These values may be shifted by the influence of terms beyond $-C_6R^{-6}$. In general, the levels lie near the top of their bins when $a_\textrm{bg}\gg\bar{a}$ and towards the bottom of the bins for $a_\textrm{bg}\ll\bar{a}$.

\begin{figure*}
\begin{center}
\includegraphics[width=2.0\columnwidth]{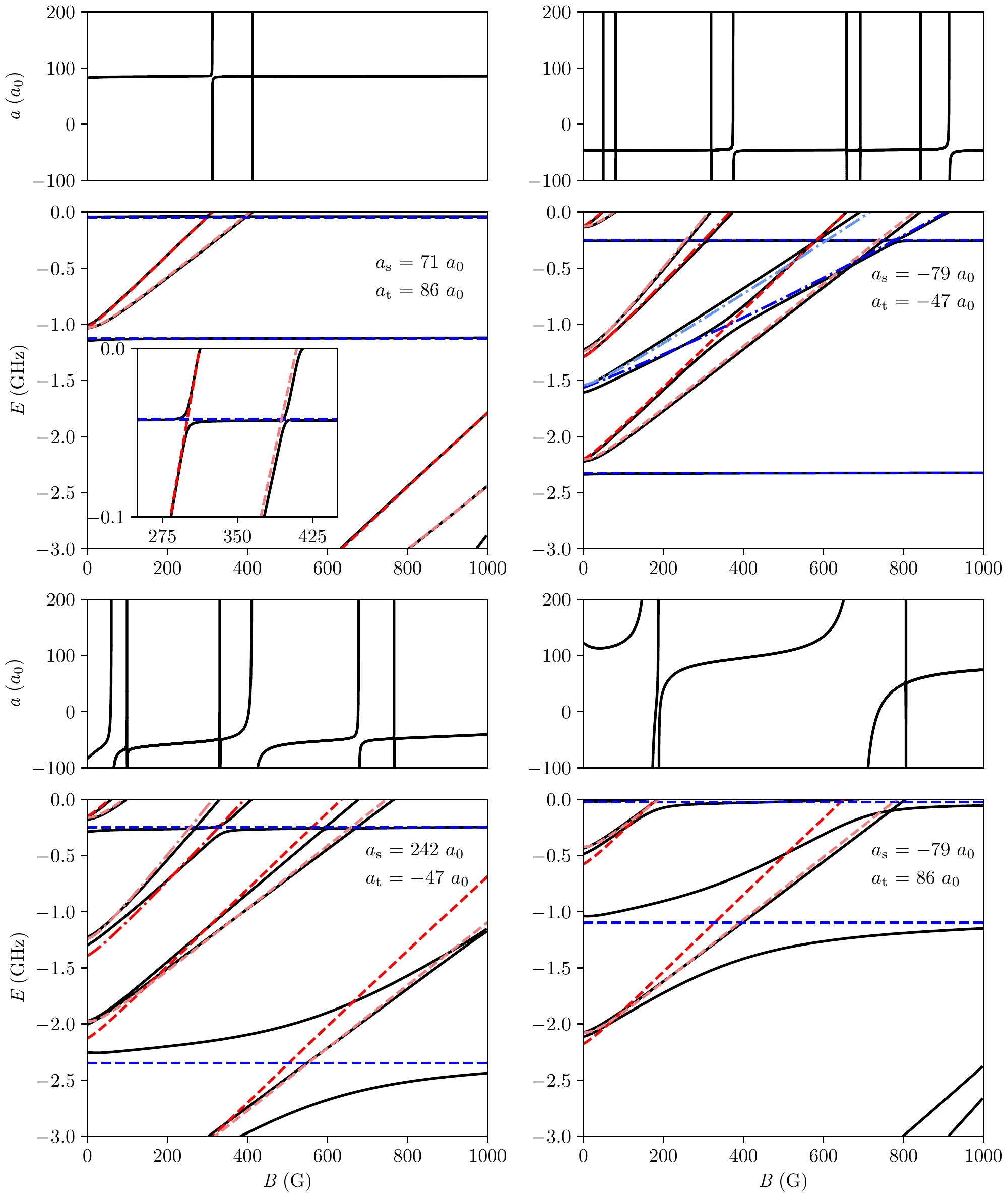}
\caption{
Near-threshold levels of Rb+CaF with $M_F=1$, neglecting anisotropy, shown relative to the energy of the lowest threshold, for four representative combinations of the singlet and triplet scattering lengths. Solid black lines show results from coupled-channel calculations. Dashed (dot-dashed) lines show uncoupled states parallel to thresholds with $f_\textrm{Rb}=1$ (2). Values of $m_{s,\textrm{CaF}}$ are encoded with red (blue) for $\frac{1}{2}$ ($-\frac{1}{2})$, with darker (lighter) colors for $m_{i,\textrm{F}}=\frac{1}{2}$ ($-\frac{1}{2}$). $m_{f,\textrm{Rb}}$ is given by $M_F-m_{s,\textrm{CaF}}-m_{i,\textrm{F}}$. Above each plot of energies is the corresponding plot of scattering length, with Feshbach resonances where states cross threshold.
\label{fig:res-iso}
}
\end{center}
\end{figure*}

\section{Bound states and resonances in the absence of anisotropy}

\subsection{Bound states below the lowest threshold}
\label{sec:crossings0}

The coupling between CaF rotational levels is fairly small at long range. It is driven mostly by the anisotropic part of the long-range interaction potential, characterized by $C_6^{(2)}$. The effects of the anisotropy will be considered in Section \ref{sec:rot}. In this section we will consider a simpler model, with the anisotropy neglected. This is expected to be a reasonably good approximation for collisions involving CaF ($n=0$), though it will neglect some additional resonances considered later.

If anisotropy is neglected, the scattering is largely controlled by the isotropic dispersion coefficient $C_6^{(0)}$ and by scattering lengths $a_\textrm{s}$ and $a_\textrm{t}$ that characterize overall phases due to the short-range parts of the singlet and triplet potentials. These scattering lengths are completely unknown for Rb+CaF, so we explore the pattern of near-threshold bound states, and the resulting Feshbach resonances, for a representative sample of values of them.

Scattering lengths take values from $-\infty$ to $+\infty$, but some values are more likely than others \cite{Gribakin:1993}. The most likely value is the mean scattering length $\bar{a}$ defined above, and for a randomly chosen potential curve that decays as $-C_6 R^{-6}$ at long range there is a 50\% probability of a scattering length between $0$ and $2\bar{a}$. To a good approximation, different interaction potentials that produce the same $a_\textrm{s}$ and $a_\textrm{t}$, and have the same value of $C_6$, have the same low-energy scattering properties and near-threshold bound states.

We use singlet and triplet potential curves based on those for Cs \cite{Berninger:Cs2:2013}, but with the value of $C_6$ replaced with $C_6^{(0)}$ for Rb+CaF. These potentials are then adjusted at short range to give the desired scattering length as described in ref.\ \cite{Berninger:Cs2:2013}. As an initial sample, we pick 3 values $a_\textrm{s}=-79$, 71 and 242~$a_0$ and $a_\textrm{t}=-47$, 86 and 297~$a_0$. These are purposely not exact multiples of $\bar{a}$, because such values can produce shape resonances at atypically low energy, and are slightly different for $a_\textrm{s}$ and $a_\textrm{t}$, because $a_\textrm{s}=a_\textrm{t}$ is a special case that produces unusually weak interchannel couplings \cite{Julienne:1997}. We consider all 9 combinations of these values of $a_\textrm{s}$ and $a_\textrm{t}$.

The solid lines in Figure \ref{fig:res-iso} shows the near-threshold energy levels for 4 combinations of $a_\textrm{s}$ and $a_\textrm{t}$, obtained from coupled-channel calculations using the package \textsc{bound}. In this case we use a basis set of fully uncoupled functions \cite{Hutson:Cs2:2008}, including only rotationless functions, $n=0$ and $L=0$. All energies are shown with respect to the (field-dependent) energy of the lowest threshold, which has approximate quantum numbers $(f_\textrm{Rb}, m_{f,\textrm{Rb}}, m_{s,\textrm{CaF}}, m_{i,\textrm{F}})=(1,1,-\frac{1}{2},\frac{1}{2})$ at fields above 50~G. All states shown have $M_F=1$, which is the same as the lowest threshold, because spin-exchange interactions cannot change $M_F$. Also shown are dashed and dot-dashed lines, parallel to thresholds but offset from them: these represent hypothetical states that would exist in the absence of interchannel couplings; the real states may be interpreted in terms of these, but with shifts and avoided crossings of various strengths due to the couplings.

The simplest case is that in Fig.\ \ref{fig:res-iso}(a), for $a_\textrm{s}= 71\ a_0$ and $a_\textrm{t}= 86\ a_0$. Here the real states lie close to the uncoupled ones, with only small shifts and weak avoided crossings. $a_\textrm{s}$ and $a_\textrm{t}$ are close to one another, so the interchannel coupling is weak, and they are comparable to $\bar{a}$, so each state lies fairly high in its bin. The near-horizontal states are those supported by the lowest threshold in the first and second bins. There is also a pair of states that originate near $-1.0$~GHz at zero field; the thresholds that support these have approximate quantum numbers $(1,0,\frac{1}{2},\frac{1}{2})$ (upper) and $(1,1,\frac{1}{2},-\frac{1}{2})$ (lower). In first order the spin-exchange coupling can change $f_\textrm{Rb}$, $m_{f,\textrm{Rb}}$ and $m_{s,\textrm{CaF}}$ by $\pm1$ while conserving $m_{f,\textrm{Rb}}+m_{s,\textrm{CaF}}$, but cannot change $m_{i,\textrm{F}}$. There is therefore a much wider avoided crossing between the near-threshold horizontal state and the upper state of the sloping pair, which is predominantly $m_{i,\textrm{F}}=\frac{1}{2}$, than with the lower one, which is predominantly $m_{i,\textrm{F}}=-\frac{1}{2}$.

A case with somewhat stronger coupling is shown in Fig.\ \ref{fig:res-iso}(b), for $a_\textrm{s}=-79\ a_0$ and $a_\textrm{t}=-47\ a_0$. Here $a_\textrm{s}$ and $a_\textrm{t}$ are negative, so the states lie much deeper in their bins than in (a). The real states still lie close to the uncoupled ones, but there is a strong avoided crossing between the states shown as red dashed and blue dot-dashed lines. States approximately parallel to the thresholds with $f_\textrm{Rb}=1$ can again be identified, with the states in the second bin now originating from around $-2.3$~GHz at zero field. These are echoed by similar states in the top bin. However, there are two further pairs of states; these are supported by thresholds with $f_\textrm{Rb}=2$, and lie in the third bin beneath their thresholds. The pair originating near $-1.6$~GHz have approximate quantum numbers $(2,1,-\frac{1}{2},\frac{1}{2})$ (lower, involving ground-state CaF but excited Rb) and $(2,2,-\frac{1}{2},-\frac{1}{2})$ (upper), while the pair originating near $-1.2$~GHz have $(2,0,\frac{1}{2},\frac{1}{2})$ (lower) and $(2,1,\frac{1}{2},-\frac{1}{2})$ (upper). Once again the strong avoided crossings are those between states with the same values of $m_{i,\textrm{F}}$.

Figs.\ \ref{fig:res-iso}(c) and (d) show further examples for cases with much stronger coupling, with $a_\textrm{s}$ and $a_\textrm{t}$ substantially different. For these cases the identification of the dashed and dotted lines is less certain, because real states are substantially shifted from the uncoupled states by interchannel couplings. Plausible assignments are shown with the same coding as in (a) and (b).

Additional bound states exist with $M_F\ne 1$. These are not connected to the lowest incoming threshold by spin-exchange coupling. However, there are additional small couplings due to the spin-spin interaction $\hat{V}^\textrm{d}$. This has matrix elements off-diagonal in $f_\textrm{Rb}$, $m_{f,\textrm{Rb}}$ and $m_{s,\textrm{CaF}}$ by $\pm1$, but can change $m_{f,\textrm{Rb}}+m_{s,\textrm{CaF}}$ by up to $\pm2$, with $M_L$ changing by up to $\mp2$ to conserve $M_\textrm{tot}=M_F+M_L$. Rotationally excited states with $L=2$ and $M_F$ from $-1$ to 3 can therefore cause additional resonances at the lowest threshold. These are expected to be narrow, and are not included in the present calculations because there is no information available on the strength of second-order spin-orbit coupling for Rb+CaF. States with other values of $L$ and $M_F$ might in principle cause resonances, but with higher-order coupling via $\hat{V}^\textrm{d}$, so the resonances will be even narrower.

\subsection{Resonances}

Each bound state with $M_F=1$ causes a magnetically tunable Feshbach resonance where it crosses threshold as a function of $B$. For all the cases considered, several such resonances exist at fields below 1000~G. However, their widths vary greatly. Figure \ref{fig:res-iso} includes a panel above each energy-level plot that shows the variation of scattering length with magnetic field. In addition, we have characterized the resonances to extract $B_\textrm{res}$, $\Delta$ and $a_\textrm{bg}$ for all resonances below 1000~G for all 9 of our representative combinations of $a_\textrm{s}$ and $a_\textrm{t}$, using the method of ref.\ \cite{Frye:resonance:2017}, and the results are given in Table \ref{tab:res-iso}.

The resonance widths may be rationalized using the same arguments about interchannel couplings used to interpret the strength of avoided crossings in Section \ref{sec:crossings0}. First, the resonances are generally broadest in cases where $a_\textrm{s}$ and $a_\textrm{t}$ are substantially different, providing strong spin-exchange coupling. Secondly, for any given combination of $a_\textrm{s}$ and $a_\textrm{t}$, the strongest resonances are those where the bound state causing the resonance has a substantial component with the same value of $m_{i,\textrm{F}}$ as the incoming channel, which for the lowest threshold is dominated by $m_{i,\textrm{F}}=\frac{1}{2}$ at fields above 50~G. The specific uncoupled states that cause the widest resonances are $(1,0,\frac{1}{2},\frac{1}{2})$ $(2,1,-\frac{1}{2},\frac{1}{2})$ and $(2,0,\frac{1}{2},\frac{1}{2})$, though in some cases their character is spread across more than one real state.

It is noteworthy that, even when $a_\textrm{s} \approx a_\textrm{t}$ and spin-exchange coupling is weak, there are resonances that are wide enough to use to control collisions or form triatomic molecules by magnetoassociation.

\begin{table*}
	\caption{Feshbach resonance positions, widths and background scattering lengths for different combinations of $a_\textrm{s}$ and $a_\textrm{t}$. The approximate quantum numbers of the uncoupled state that causes the resonance are given in each case.
Asterisks indicate cases where this uncoupled state is substantially mixed with the least-bound state in the incoming channel where it crosses threshold.}
\begin{tabular}{r r r c r r r r r r r}
\hline\hline
	$a_\textrm{s}\ (a_0)$ & $a_\textrm{t}\ (a_0)$ & $B_\textrm{res}$ (G) & $\Delta$ (G) & $a_\textrm{bg}\ (a_0)$ & $f_\textrm{Rb}$ &  $m_{f,\textrm{Rb}}$ &  $m_{s,\textrm{CaF}}$ & $m_{i,\textrm{F}}$ &  $\eta$ &\\
\hline

	$-79$ & $-47$ & 50 & $-6.8\times10^{-4}$ & $-47$ & 1 & 0 & $\frac{1}{2}$ & $\frac{1}{2}$ & $-1$\\
		    & & 81 & $-2.3\times10^{-4}$ & $-47$ & 1 & 1 & $\frac{1}{2}$ & $-\frac{1}{2}$ & $-1$\\
		    & & 319 & $-8.2\times10^{-3}$ & $-46$ & 2 & 1 & $\frac{1}{2}$ & $-\frac{1}{2}$  & $-3$\\
		    & & 375 & $-4.8\times10^{-1}$ & $-46$ & 2 & 0 & $\frac{1}{2}$ & $\frac{1}{2}$ & $-3$\\
		    & & 658 & $-1.7\times10^{-2}$ & $-46$ & 1 & 0 & $\frac{1}{2}$ & $\frac{1}{2}$ & $-2$\\
		    & & 692 & $-1.0\times10^{-2}$ & $-46$ & 2 & 2 & $-\frac{1}{2}$ & $-\frac{1}{2}$ &$-3$\\
		    & & 843 & $-8.6\times10^{-4}$ & $-45$ & 1 & 1 & $\frac{1}{2}$ & $-\frac{1}{2}$ & $-2$\\
		    & & 914 & $-1.1$ & $-46$ & 2 & 1 & $-\frac{1}{2}$ & $\frac{1}{2}$ & $-3$ \\[0.2cm]
	
	$-79$ & 86 & 164 & 16 & 112 & 1 & 0 & $\frac{1}{2}$ & $\frac{1}{2}$ & $-2$ &* \\
		 & & 188 & 3.3 & 27 & 1 & 1    & $\frac{1}{2}$ & $-\frac{1}{2}$ & $-2$\\
		 & & 688 & 49 & 87 & 1 & 0 & $\frac{1}{2}$ & $\frac{1}{2}$ & $-3$ &*\\
		 & & 806 & $3.0\times10^{-2}$ & 51 & 1 & 1 & $\frac{1}{2}$ & $-\frac{1}{2}$ & $-3$\\[0.2cm]
	
	$-79$ & 297 &124 & 1.8 & 273 & 1 & 1 & $\frac{1}{2}$ & $-\frac{1}{2}$ & $-1$ \\
		  & & 599 & 88 & 383 & 1 & 0 & $\frac{1}{2}$ & $\frac{1}{2}$ & $-2$ &*\\
		  & & 725 & $5.7\times10^{-2}$ & 117 & 1 & 1 & $\frac{1}{2}$ & $-\frac{1}{2}$ & $-2$ \\
		  & & 934 & $9.5\times10^{-2}$ & 294 & 2 & 1 & $\frac{1}{2}$ & $-\frac{1}{2}$ & $-3$ \\
		  & & 953 & $3.8\times10^{-1}$ & 292 & 2 & 0 & $\frac{1}{2}$ & $\frac{1}{2}$ & $-3$ \\[0.2cm]
	
	71 & $-47$ & 99 & $-32$ & $-83$ & 1 & 0 & $\frac{1}{2}$ & $\frac{1}{2}$ & $-1$ \\
		 & & 134 & $-2.4\times10^{-1}$ & $-144$ & 1 & 1 & $\frac{1}{2}$ & $-\frac{1}{2}$ & $-1$ \\
		 & & 343 & $-1.9\times10^{-1}$ & $-49$ & 2 & 1 & $\frac{1}{2}$ & $-\frac{1}{2}$ & $-2$ \\
		 & & 455 & $-27$ & $-51$ & 2 & 0 & $\frac{1}{2}$ & $\frac{1}{2}$ & $-2$\\
		 & & 689 & $-3.2$ & $-50$ & 1 & 0 & $\frac{1}{2}$ & $\frac{1}{2}$ & $-2$ \\
		 & & 776 & $-3.9\times10^{-2}$ & $-49$ & 1 & 1 & $\frac{1}{2}$ & $-\frac{1}{2}$ & $-2$\\[0.2cm]
	
	71 & 86 & 312  & $1.1\times10^{-1}$ & 85 & 1 & 0 & $\frac{1}{2}$ &  $\frac{1}{2}$ & $-2$ \\
		& & 413 & $9.4\times10^{-4}$ & 85 & 1 &  1 & $\frac{1}{2}$ & $-\frac{1}{2}$ & $-2$ \\[0.2cm]
	
	71 & 297 & 172 & 18 & 202 & 1 & 0 & $\frac{1}{2}$ & $\frac{1}{2}$ & $-2$ &*\\
	       & & 285 & $9.5\times10^{-2}$ & 185 & 1 & 1 & $\frac{1}{2}$ & $-\frac{1}{2}$ & $-2$ \\
	       & & 860 & $2.9\times10^{-2}$ & 279 & 2  & 1 & $\frac{1}{2}$ & $-\frac{1}{2}$ & $-3$ \\
	       & & 952 & $3.2\times10^{-1}$ & 294 & 2 & 0 & $\frac{1}{2}$ & $\frac{1}{2}$ & $-3$ \\[0.2cm]
	
	242 & $-47$ & 61 & $-2.8$ & $-65$ & 1 & 0 & $\frac{1}{2}$ & $\frac{1}{2}$ & $-1$ \\
		  & & 99 & $-1.8\times10^{-1}$ & $-67$ & 1 & 1 & $\frac{1}{2}$ & $-\frac{1}{2} $ & $-1$ \\
		  & & 331 & $-9.6\times10^{-2}$ & $-50$ & 2 &  1 & $\frac{1}{2}$ & $-\frac{1}{2}$ & $-3$ \\
		  & & 413 & $-10$ & $-54$ & 2 & 0 & $\frac{1}{2}$ & $\frac{1}{2}$ & $-3$\\
		  & & 678 & $-1.4$ & $-50$ & 1 & 0 & $\frac{1}{2}$ & $\frac{1}{2}$ & $-2$\\
		  & & 766 & $-4.4\times10^{-2}$ & $-48$ & 1 &  1 & $\frac{1}{2}$ & $-\frac{1}{2}$ & $-2$ \\[0.2cm]
	
	242 & 86 & 248 & 4.0 & 93 & 1 & 0 & $\frac{1}{2}$ & $\frac{1}{2}$ & $-2$ &*\\
	    &    & 302 & $1.4\times10^{-1}$ & 86 & 1 & 1 & $\frac{1}{2}$ & $-\frac{1}{2}$ & $-2$ \\[0.2cm]
	
	242 & 297 & 134 & $3.3\times10^{-3}$ & 300 & 1 & 0 & $\frac{1}{2}$ & $\frac{1}{2}$ & $-2$ \\
		& & 187 & $1.9\times10^{-4}$ & 300 & 1 & 1 & $\frac{1}{2}$ & $-\frac{1}{2}$ & $-2$\\
		& & 838 & $7.8\times10^{-4}$ & 302 & 2 & 1 & $\frac{1}{2}$ & $-\frac{1}{2}$ & $-3$\\
		& & 951 & $1.3\times10^{-1}$ & 301 & 2 & 0 & $\frac{1}{2}$ & $\frac{1}{2}$ & $-3$\\
		& & 990 & $1.8\times10^{-4}$ & 300 & 1 & 0 & $\frac{1}{2}$ & $\frac{1}{2}$ & $-3$\\[0.1cm]
\hline\hline
\end{tabular}
\label{tab:res-iso}
\end{table*}

\section{The role of C\lowercase{a}F rotation}
\label{sec:rot}

There can also be resonances due to states supported by rotationally excited thresholds. This section will consider the structure of such states and the likelihood that they produce resonances at experimentally accessible fields.

The thresholds for CaF ($n=1$) are from 20 to 30~GHz above the lowest threshold, so states that can cause Feshbach resonances must be bound by about this amount. The outer turning point at this depth is at around $R=30\ a_0$. The potential anisotropy at this distance, due to dispersion and induction, is around 1~GHz. This is substantially less than the CaF rotational spacing, so will cause only weak mixing between different CaF rotational states at this distance. However, it is substantially larger than the rotational constant of the triatomic complex, $B=\hbar^2/(2\mu R^2)$, which is about 60~MHz  at this distance. It is also larger than the spin-rotation coupling constant, $\gamma\approx 40$~MHz. The long-range anisotropy is thus sufficient to quantize $n$ along the intermolecular axis, with projection $K$, instead of along the axis of the field. This is exactly analogous to the situation for Van der Waals complexes in coupling case 2 \cite{Hutson:AMVCD:1991}.

For each CaF rotational level $(n,K)$ there will be a set of spin states, labeled at fields above 50~G by $(f_\textrm{Rb}, m_{f,\textrm{Rb}}, m_{s,\textrm{CaF}}, m_{i,\textrm{F}})$. Each such set $(n,K)$ will sample the short-range singlet and triplet potentials over a different range of Jacobi angles $\theta$, so each group will be characterized by different singlet and triplet scattering lengths $a_\textrm{s}(n,K)$ and $a_\textrm{t}(n,K)$. These will probably be unrelated to the corresponding quantities for the channels with $n=0$, $a_\textrm{s}(0,0)$ and $a_\textrm{t}(0,0)$ (designated simply $a_\textrm{s}$ and $a_\textrm{t}$ in Sec.\ \ref{sec:crossings0}). For a particular interaction potential, the sets of spin states for $n>0$ may therefore lie at quite different depths within their bins from those for $n=0$. The patterns of levels will nevertheless be characterized by $a_\textrm{s}(n,K)$ and $a_\textrm{t}(n,K)$ and by quantum numbers $(f_\textrm{Rb}, m_{f,\textrm{Rb}}, m_{s,\textrm{CaF}}, m_{i,\textrm{F}})$, in a similar way to those for the states with $n=0$ described above.

For Rb+CaF, the spin-exchange interaction may be characterized in terms of an anisotropic surface $V^-(R,\theta)$ that is half the difference between the singlet and triplet surfaces,
\begin{equation}
V^-(R,\theta) = \textstyle{\frac{1}{2}} \left[V^0(R,\theta)-V^1(R,\theta)\right].
\end{equation}
This may be expanded in Legendre polynomials,
\begin{equation}
V^-(R,\theta) = \sum_\lambda V^-_\lambda(R) P_\lambda(\cos\theta).
\end{equation}
Such a potential is diagonal in $K$, but each term in the expansion can couple $(n,K)=(0,0)$ to $(\lambda,0)$. The term $V^-_1(R)$ can thus couple an incoming state at the lowest threshold to states with $(n,K)=(1,0)$. The spin selection rules are the same as for $n=0$, so the strongest resonances will be those due to states dominated by $m_{i,\textrm{F}}=\frac{1}{2}$. As for $n=0$, there are 3 such uncoupled states, with quantum numbers $(f_\textrm{Rb}, m_{f,\textrm{Rb}}, m_{s,\textrm{CaF}}, m_{i,\textrm{F}})=(1,0,\frac{1}{2},\frac{1}{2})$ $(2,1,-\frac{1}{2},\frac{1}{2})$ and $(2,0,\frac{1}{2},\frac{1}{2})$. $V^-(R,\theta)$ is strongly anisotropic at short range, so there will always be some intermolecular distance $R$ where it matches the separation between the incoming and resonant thresholds, where nonadiabatic couplings can occur by extension of the theory of ref.\ \cite{Julienne:1997}.

For a potential with long-range form $-C_6 R^{-6}$, the binding energy of a state that lies below the top bin is approximately proportional to $(|\eta|+\frac{1}{8})^3$ \cite{LeRoy:1970, Gribakin:1993}. Here $\eta$ is a noninteger vibrational quantum number, with integer values at the bin boundaries. For an unknown potential of sufficient depth, the fractional part of $\eta$ may be regarded as a uniform random variable. Since there is one state in each bin, this allows calculation of the probability that there is a state within any particular range of energies.

As seen in Sec.\ \ref{sec:crossings0}, the states that can cause strong resonances traverse about 3~GHz of binding energy between zero field and 1000~G. Since the thresholds with $(n,f_\textrm{Rb})=(1,1)$ lie about 20~GHz above $(0,1)$, the zero-field binding energy of a state must be between 20 and 23~GHz if it is to cause a Feshbach resonance below 1000~G. For an unknown potential, there is only about a 19\% probability that there is a state with a binding energy in this range.
The corresponding probability for $n=2$ is about 9\%, and the probabilities decrease for successively higher $n$, because the bins are correspondingly wider at the required binding energy.

The overall conclusion of this section is that there \emph{may} be resonances due to states involving rotationally excited CaF, but that they will occur at fields below 1000~G for a fairly small subset of possible interaction potentials. In any case, the mixing between rotational states of CaF due to long-range anisotropy is weak enough that it will not affect the likelihood of resonances due to the ground rotational state.

\section{Potential effects of chaos}
\label{sec:chaos}

The interaction potentials for Rb+CaF are very strongly anisotropic at short range, and provide strong coupling between CaF rotational and vibrational states. It is quite likely that Rb+CaF will possess short-range states that exhibit quantum chaos, in the same way as alkali-metal 3-atom \cite{Mayle:2012, Frye:triatomic-complexes:2021} and 4-atom systems \cite{Mayle:2013, Christianen:density:2019}. The onset of chaos has also been studied in Li+CaH and Li+CaF \cite{Frye:chaos:2016}.

For Rb+CaF, the density of short-range singlet vibrational states at the energy of the lowest threshold has been estimated as 4~K$^{-1}$ \cite{Jurgilas:magnetic:2021}, corresponding to a mean spacing of 5~GHz. If these states are fully chaotic, it is likely to produce structure in the singlet scattering length on this energy scale. However, the hyperfine couplings in singlet states will be small, probably dominated by nuclear electric quadrupole couplings of no more that a few MHz, which is tiny compared to the state separations. Furthermore, Zeeman shifts are very small for singlet states. At most, the presence of chaos at short range might make the singlet scattering length different for collisions involving Rb($f=1$) and Rb($f=2$). This would affect the details of the level structure, but not the probabilities of observing Feshbach resonances.

The density of short-range triplet states at threshold is likely to be much smaller, perhaps by an order of magnitude. This corresponds to a mean spacing of order 50~GHz. The difference arises because the density of states for an atom-diatom system scales approximately with $D^{3/2}$ \cite{Frye:triatomic-complexes:2021}, where $D$ is the well depth, and the triplet surface of Rb+CaF is expected to be substantially shallower than the singlet surface, as for the alkali-metal dimers. The hyperfine couplings for triplet states will be comparable to those for the separated atom and molecule (6.8 GHz for Rb, 120 MHz for CaF) but these are still substantially smaller than the likely spacings between short-range triplet states. Zeeman effects are also much larger for triplet states than for singlet states, but are still only a few GHz at fields below 1000~G, so will not cause substantial mixings between short-range triplet states.

It thus appears that the qualitative arguments in this paper about the patterns of energy levels and likelihood of Feshbach resonances will remain valid even if the short-range levels of Rb+CaF exhibit quantum chaos.

\section{Conclusions}
\label{sec:conc}
We have investigated magnetically tunable Feshbach resonances that may be expected in collisions between molecules in $^2\Sigma$ states and alkali-metal atoms, focussing on the prototype system Rb+CaF. The details of the short-range interaction potential are unknown, but expected to have minor influence, except to determine singlet and triplet scattering lengths $a_\textrm{s}$ and $a_\textrm{t}$. We have carried out coupled-channel calculations of the near-threshold bound states and scattering properties for a variety of values of these scattering lengths. We find that the large majority of plausible interaction potentials produce multiple resonances at magnetic field below 1000~G, which are likely to be experimentally accessible. In each case, at least some of these resonances are wide enough to be experimentally useful for tuning scattering lengths or for forming triatomic molecules by magnetoassociation.

The patterns of bound states may be understood in terms of underlying uncoupled states that lie parallel to atom-molecule thresholds as a function of magnetic field. There are varying degrees of coupling between these states, which depend on the values of $a_\textrm{s}$ and $a_\textrm{t}$. The coupling is weakest when $a_\textrm{s}$ and $a_\textrm{t}$ are similar. The widths of the resonances may be explained in terms of the nature of the states that cross threshold, together with effects due to the scattering lengths.

We have considered the effect of potential anisotropy, which causes coupling between CaF rotational states. This coupling is very strong at short range. Even at long range, it is sufficient to quantize the CaF rotation along the intermolecular axis instead of along the magnetic field. It is likely that each rotational state of CaF will be characterized by different values of the singlet and triplet scattering lengths. We have found that there is a small but significant probability of additional wide resonances due to states supported by rotationally excited thresholds. We have also considered the potential influence of chaotic behavior for short-range states of Rb+CaF. We expect that, even if present, it will have limited effects on the long-range states that are principally responsible for the resonances and will not change the qualitative conclusions.

This work indicates that atom-molecule systems such as Rb+CaF will have a rich spectrum of magnetically tunable Feshbach resonances at experimentally accessible magnetic fields. The resonances can be used to form a more detailed understanding of the atom-diatom potential energy surfaces. Much new physics will be accessible when these resonances are located. For example, a resonance can be used to tune the s-wave scattering length for interspecies collisions. In this way we can expect to find favorable conditions for sympathetic cooling, which can greatly increase the phase-space density of the molecular gas. The resonances may also be used to form polyatomic molecules by magnetoassociation. Many applications have already been identified for such molecules. They have unique advantages for probing interactions beyond the Standard Model that violate time-reversal symmetry \cite{Kozyryev:2017, Hutzler:2020} and for testing theories of ultralight dark matter \cite{Kozyryev:2021}. Their usefulness for quantum information processing has been highlighted \cite{Wei:2011, Yu:2019}, and the very large number of stable, accessible internal states make them interesting as qudits \cite{Sawant:qudit:2020}. They can also be used to explore a rich diversity of many-body phenomena such as quantum magnetism \cite{Wall:2015}.

\section*{Rights retention statement}

For the purpose of open access, the authors have applied a Creative Commons Attribution (CC BY) licence to any Author Accepted Manuscript version arising from this submission.

\section*{Data availability statement}

The data presented in this work are available from Durham University \cite{DOI_data-RbCaF-resonances}.

\section*{Acknowledgement}
We are grateful to Matthew Frye for valuable discussions and to Ruth Le Sueur for software support. This work was supported by the U.K. Engineering and Physical Sciences Research Council (EPSRC) Grant Nos.\ EP/M507854/1, EP/P01058X/1, EP/T518001/1, EP/W00299X/1, EP/V011499/1 and EP/V011677/1.

\bibliographystyle{../long_bib}
\bibliography{../all,Caf+Rb-resonances-data}
\end{document}